\documentclass[aps,prd,twocolumn,a4paper]{revtex4-1}
\usepackage{ulem}
\usepackage{color}
\usepackage{graphicx}
\usepackage{hyperref}   
\usepackage{appendix}
\usepackage{epsfig}
\usepackage{epstopdf}
\usepackage{amsmath}
\usepackage{subfig}
\usepackage{comment}
\usepackage[dvipsnames]{xcolor}
\newcommand{\be}{\begin{equation}}
\newcommand{\ee}{\end{equation}}
\newcommand{\ba}{\begin{eqnarray}}
\newcommand{\ea}{\end{eqnarray}}

\begin{document}

\title{Heavy quark dynamics in a hot magnetized  QCD medium}

\author{Manu Kurian $^{a, b}$ } 
\email{manu.kurian@iitgn.ac.in}
\author{Santosh K. Das $^{c}$}
\email{dsantoshphy@gmail.com}
\author{Vinod Chandra $^{a}$}
\email{vchandra@iitgn.ac.in}
\affiliation{$^{a}$Indian Institute of Technology Gandhinagar, Gandhinagar-382355, Gujarat, India  \\
$^{b}$Department of Physics, McGill University, 3600 University Street, Montreal, QC, H3A 2T8, Canada\\ $^{c}$School of Physical Science, Indian Institute of Technology Goa, Ponda-403401, Goa, India}


\begin{abstract}
The heavy quark drag and momentum diffusion have been investigated in a hot magnetized 
quark-gluon plasma, along the directions parallel and perpendicular to the magnetic field. The 
analysis is done within the framework of Fokker-Planck dynamics by considering the heavy 
quark scattering with thermal quarks and gluons at the leading order in the coupling constant. 
An extended quasiparticle model is adopted to encode the thermal QCD medium interactions 
in the presence of a magnetic field. Further, the higher Landau level effects on the temperature 
behaviour of the parallel and perpendicular components of the drag force and diffusion coefficients 
have studied. It has been observed that both the equation of state and the magnetic field play key 
roles in the temperature dependence of the heavy quark dynamics.

\end{abstract}



\maketitle

 \section{Introduction}
 
The heavy ion-collision experiments at Relativistic Heavy 
Ion Collider (RHIC) and at the Large Hadron Collider (LHC) set the stage to investigate the 
deconfined state of the nuclear matter called quark-gluon plasma (QGP), as a near-ideal
fluid~\cite{STAR, Aamodt:2010pb,Heinz:2008tv}. Recent studies revealed the presence of 
ultra-intense magnetic field in the non-central asymmetric collisions~\cite{Skokov:2009qp,
Zhong:2014cda,deng,Roy:2017yvg}.  Inclusion of the magnetic field to the theoretical investigations regarding the QGP/hot QCD matter is seen to  affect  its  transport and 
thermodynamic properties in a significant way~\cite{Karmakar:2019tdp,Bandyopadhyay:2017cle,Koothottil:2018akg,
Hattori:2017qih,Hattori:2016cnt,Kurian:2018dbn,Rath:2019vvi,Mohanty:2018eja,Das:2019wjg,Ghosh:2018cxb}.
 In particular, the novel phenomena 
such as chiral magnetic effect~\cite{Fukushima:2008xe,Sadofyev:2010pr,She:2017icp}, 
chiral vortical effect~\cite{Kharzeev:2015znc,Avkhadiev:2017fxj,Yamamoto:2017uul}, 
chiral charge separation~\cite{Huang:2015fqj}, magnetic catalysis~\cite{Gusynin:1995nb}, 
and more recently the realization of global $\Lambda-$hyperon polarization in the
RHIC~\cite{STAR:2017ckg,Becattini:2016gvu} opens up new directions in the investigation on magnetized QGP.

Heavy quarks (HQs), mainly charm and bottom, are identified as effective probes to characterize 
the properties of the QGP~\cite{Dong:2019unq, Prino:2016cni, Rapp:2018qla,Cao:2018ews, 
Aarts:2016hap,Andronic:2015wma}. The HQs are mostly created in the early stages of collision 
and  propagate through the bulk medium (QGP) while interacting with its constituents (light quarks and gluons). 
The HQs are the witness of  the entire space-time evolution of the 
bulk medium. There have been several attempts to investigate the  HQ dynamics in 
QCD matter~\cite{Svetitsky:1987gq,GolamMustafa:1997id,Moore:2004tg,vanHees:2005wb,
vanHees:2007me,Gossiaux:2008jv,Das:2009vy,Alberico:2013bza,Uphoff:2012gb,Young:2011ug,Cao:2013ita,
Das:2013kea,Das:2015ana,Song:2015sfa,Cao:2016gvr,Scardina:2017ipo,Singh:2018wps} 
and the related experimental observables such as nuclear suppression factor, elliptic flow and heavy baryon to meson 
ratio which serve as direct QGP probes~\cite{Adare:2006nq,Adler:2005xv,Acharya:2018ckj}. 

Recently, it has been recognized that the strong electromagnetic fields created at early times  
of the heavy-ion collisions can affect the HQs dynamics. HQ directed flow 
($v_1$)~\cite{Das:2016cwd} is identified as a potential probe of the strong initial electromagnetic 
field created in heavy-ion collisions. This transient field can induce opposite $v_1$ 
for the charm and anti-charm quarks due to their opposite charge. The $v_1$ for hadrons containing HQs is 
predicted to be several orders of magnitude larger than that for hadrons containing light quark; a prediction 
that appears to be vindicated by early experimental results at both RHIC and LHC energies~\cite{Grosa:2018zix,
Singha:2018cdj,Adam:2019wnk}. However, recent calculations~\cite{Das:2016cwd, Chatterjee:2018lsx, Coci:2019nyr} 
on the HQ directed flow due to the electromagnetic field, within the Langevin dynamics, ignore the impact 
of the magnetic field on HQ transport coefficients. Hence, it is an interesting aspect to investigate the 
HQ drag and momentum diffusion in the presence of the magnetic field and explore its consequences on 
experimental observables.

The light quarks/antiquarks degree of freedom are affected by the magnetic field and follow the Landau 
Level dynamics in the thermal equilibrium. Whereas the electrically neutral gluons indirectly coupled 
to the magnetic field through the self-energy via quark/antiquark loop. 
The transport coefficients of the magnetized QGP have been studied in the lowest Landau level (LLL)
approximation~\cite{Hattori:2017qih,Kurian:2017yxj} and the recent studies~\cite{Fukushima:2017lvb,
Kurian:2018qwb,Kurian:2019fty} revealed the significance of higher Landau levels (HLLs) in the analysis. 
Several investigations on the static properties of HQs and quarkonia in the presence of magnetic field 
have been done in Refs.~\cite{Machado:2013rta,Bonati:2015dka,Hasan:2017fmf,Singh:2017nfa,
Guo:2015nsa,Gubler:2015qok}. There are a few attempts  within  the holographic and conformal 
field theory description of the HQ transport with a strong magnetic field background~\cite{Finazzo:2016mhm,
Rajagopal:2015roa,Kiritsis:2011ha}. Since the external magnetic field constraints the light quarks/antiquarks 
motion in preferred spatial direction (either parallel or antiparallel to the field), one needs to analyze the HQ 
dynamics in both parallel and perpendicular to the magnetic field. In the recent work~\cite{Fukushima:2015wck}, 
the authors have showed that the diffusion coefficient of HQ became anisotropic in the presence 
of a strong magnetic field in the LLL approximation. 

The present article primarily focuses on the study of longitudinal and transverse components of the 
momentum diffusion and drag of the HQ in the magnetized QGP, incorporating the effects of hot 
QCD medium interactions and HLLs. To that end, the modeling of the local momentum distribution 
function of gluons and quarks are essential such that the realistic equation of state (EoS) of the QGP 
could be mimicked. At this juncture, a recently proposed effective fugacity quasiparticle model 
(EQPM)~\cite{Chandra:2011en,Chandra:2007ca} has been employed in the analysis. The thermal 
QCD medium effects are reflected in the temperature dependence of the effective fugacity of the EQPM 
phase space distribution function. The quasiparticle description of 
HQ dynamics in the QGP has been investigated in the absence of the magnetic field in Refs.~\cite{Das:2012ck,Chandra:2015gma}. 

For the quantitative description of the HQ dynamics in the magnetized QGP, the relativistic Boltzmann 
equation needs to be solved by embedding the proper collision integral in the presence of external 
magnetic field background. The motion of HQ can be analyzed as a Brownian motion while considering 
their perturbative interaction, and the large HQ mass allows to assume for the low momentum transfer 
scattering~\cite{Svetitsky:1987gq}. Under such constraints, the relativistic transport equation can be 
reduced to the Fokker-Planck equation. This assumption has been widely employed in the investigation 
of HQ propagation in the medium~\cite{Moore:2004tg,Das:2013kea,GolamMustafa:1997id}. The current 
investigation is done within the framework of the Fokker-Planck dynamics by analyzing the scattering of 
HQs with thermal gluons and quarks separately in the presence of the magnetic field. At the leading 
order in the coupling constant $\alpha_s$, the HQ scattering with gluons and quarks are mediated by 
a one-gluon exchange. The HLLs contribution to the HQ dynamics in the 
magnetized QGP has also been estimated.

The article is organized as follows. Section II is devoted to the mathematical formulation of the gluonic 
and quark contributions to the HQ drag and diffusion within the framework of the Fokker-Planck equation 
followed by the quasiparticle modeling of the magnetized QGP. The discussions on the effect of thermal 
QCD medium and HLLs on the temperature behaviour of the HQ dynamics are presented in Section III. 
Section VI contains the conclusions and outlook of the article.

\section{Heavy quark drag and diffusion in magnetized QGP }

HQs are subjected to random motion at the finite temperature due to the scattering with thermally excited 
quarks and gluons. Since the kinematics of quarks and gluons are different in the presence of the magnetic 
field, the HQ scattering with quarks and gluons need to be calculated separately~\cite{Hattori:2018yqo}. 
Note that the mass of HQ, $M_{HQ}$, is assumed to follow $M_{HQ}\gg \sqrt{q_feB}$, where $q_fe$ is the 
fractional charge of the quark of flavor $f$, such that HQ motion is not directly affected by the magnetic field. 
In the Ref.~\cite{Das:2016cwd}, the authors have included the Lorentz force in the analysis of the direct flow 
of the charm quark. The HQ dynamics 
can be understood in terms of the phase space distribution function with the prescription of transport theory.

\subsection{Thermal gluon contribution to HQ transport}

In this section, we are considering the gluonic contribution to the HQ dynamics while travelling in the QGP 
in the presence of the strong magnetic field ${\bf{B}}=B\hat{z}$. Note that the magnetic field affects 
the gluon dynamics through the self-energy and the Debye screening 
mass in the system~\cite{Kurian:2018dbn}. The dynamics of HQ can be understood in terms of the drag 
and momentum diffusion in the medium.

\subsubsection*{Formalism of HQ drag and diffusion} 

The evolution of the HQ momentum distribution function $f_{HQ}$ in the QGP can be described by the 
Boltzmann equation as~\cite{Svetitsky:1987gq},
\begin{equation}\label{1.1}
p^{\mu}\partial_{\mu}f_{HQ}=\bigg(\dfrac{\partial f_{HQ}}{\partial t}\bigg)_{col}.
\end{equation}
 The term $(\frac{\partial f_{HQ}}{\partial t})_{col}$ is the relativistic collision integral that quantifies 
 the rate of change in the HQ distribution function due to the interactions with thermal gluons in the 
 medium. For the two-body collision, the collision term takes the following form~\cite{Svetitsky:1987gq},
\begin{align}\label{1.2}
\bigg(\dfrac{\partial f_{HQ}}{\partial t}\bigg)_{col}=&\int{d^3{\bf q}\bigg[\omega({\bf p}+{\bf q},{\bf q})f_{HQ}({\bf p}+{\bf q})}\nonumber\\
&-\omega({\bf p},{\bf q})f_{HQ}({\bf p})\bigg].
\end{align}
The quantity $\omega({\bf p},{\bf q})$ is the rate of collisions with gluons that 
change the HQ momentum from ${\bf p}$ to ${\bf p}-{\bf q}$ and defined as,
\begin{align}\label{1.3}
\omega({\bf p},{\bf q})=\int{\dfrac{d^3{\bf k}}{(2\pi)^3}f_{g}({\bf k})v\sigma_{{\bf p},{\bf k}\rightarrow {\bf p}-{\bf q},{\bf k}+{\bf q}}},
\end{align}
where the interaction cross-section $\sigma_{{\bf p},{\bf k}\rightarrow {\bf p}-{\bf q},{\bf k}+{\bf q}}$ is 
related to the matrix element $\mathcal{M}_{HQ,g}$ of the HQ scattering process with 
gluons. Here, $f_{g}({\bf k})$ is the momentum distribution of gluons and $v$ is the 
relative velocity between the colliding particles. Note that in the integrand of Eq.~(\ref{1.2}), the 
first term constitutes the gain term through the scattering whereas the second term represents 
the loss out of the volume element around the HQ momentum ${\bf p}$. 

The integral operator in the Boltzmann equation can be simplified by employing the Landau approximation 
which assumes that most of the HQ-qluon scattering is soft with small momentum transfer. 
Hence, we can expand $\omega({\bf p}+{\bf q},{\bf q})~f_{HQ}({\bf p}+{\bf q})$ up to the 
second order of the momentum transfer ${\bf q}$ as,
\begin{align}\label{1.4}
\omega({\bf p}+{\bf q},{\bf q})~f_{HQ}({\bf p}+{\bf q})\approx & \omega({\bf p},{\bf q})~f_{HQ}({\bf p})
+ {\bf q}\cdot\dfrac{\partial}{\partial {\bf p}}\big[\omega f_{HQ}\big]\nonumber\\
&+ \dfrac{1}{2}q_iq_j\dfrac{\partial^2}{\partial p_i\partial p_j}\big[\omega f_{HQ}\big].
\end{align}
The relativistic Boltzmann equation can be reduced to the Fokker-Planck equation by 
employing the Eqs.~(\ref{1.2}),~(\ref{1.3}) and~(\ref{1.4}) in the Eq.~(\ref{1.1}) and takes the form as follows,
\begin{align}\label{1.5}
\dfrac{\partial f_{HQ}}{\partial t}=\dfrac{\partial}{\partial p_i}\bigg[A_i({\bf p})~f_{HQ}
+\dfrac{\partial}{\partial p_j}\big[B_{ij}({\bf p})~f_{HQ}\big]\bigg],
\end{align}
where $A_i$ and $B_{ij}$ measure the drag force and momentum diffusion of the 
HQs, respectively. For the process, $HQ(p)+g(k)\rightarrow HQ(p^{'})+g(k^{'})$, 
where $g$ stands for gluons in the magnetized thermal medium, the HQ drag 
and momentum diffusion takes the following forms~\cite{Svetitsky:1987gq,GolamMustafa:1997id},
\begin{align}\label{1.6}
A_i=&\dfrac{1}{d_{HQ}}\dfrac{1}{2E_p}\int{\dfrac{d^3{\bf k}}{(2\pi)^3E_k}}\int{\dfrac{d^3 {\bf p}^{'}}{(2\pi)^3E_{p^{'}}}}\int{\dfrac{d^3{\bf k}^{'}}{(2\pi)^3E_{k^{'}}}}\nonumber\\
&\mid\mathcal{M}_{HQ,g}\mid^2(2\pi)^4\delta^4(p+k-p^{'}-k^{'})f_g({\bf k})\nonumber\\
&\Big(1+f_g({\bf k}^{'})\Big)\big({\bf p}-{\bf p}^{'}\big)_i \nonumber\\
&\equiv <<\big({\bf p}-{\bf p}^{'}\big)_i>>,
\end{align}
and
\begin{align}\label{1.7}
B_{ij}=&\dfrac{1}{2d_{HQ}}\dfrac{1}{2E_p}\int{\dfrac{d^3{\bf k}}{(2\pi)^3E_k}}\int{\dfrac{d^3{\bf p}^{'}}{(2\pi)^3E_{p^{'}}}}\int{\dfrac{d^3{\bf k}^{'}}{(2\pi)^3E_{k^{'}}}}\nonumber\\
&\mid\mathcal{M}_{HQ,g}\mid^2(2\pi)^4\delta^4(p+k-p^{'}-k^{'})f_g({\bf k})\nonumber\\
&\Big(1+f_g({\bf k}^{'})\Big)({\bf p}-{\bf p}^{'})_i ({\bf p}-{\bf p}^{'})_j\nonumber\\
&\equiv <<\big({\bf p}-{\bf p}^{'}\big)_i\big({\bf p}-{\bf p}^{'}\big)_j>>,
\end{align}
respectively. Here, $d_{HQ}$ is the statistical degeneracy of the HQ. The Eq.~(\ref{1.6}) indicates 
that the HQ drag is the measure of the thermal average of the momentum transfer ${\bf q}={\bf p}-{\bf p}^{'}$ 
due to the scattering of HQ with the thermal particles. Whereas the momentum diffusion in Eq.~(\ref{1.7}) 
measures the thermal average of square of the momentum transfer. In the static limit ${\bf p}\rightarrow 0$, 
we can consider $B_{ij}\rightarrow K\delta_{ij}$~\cite{Svetitsky:1987gq} with $K$ is the HQ diffusion coefficient. 
The magnetic field provides the 
preferred spatial direction and we need to consider the HQ motion parallel and perpendicular to the 
magnetic field. The longitudinal and transverse components of HQ drag and diffusion in the presence 
of magnetic field (${\bf{B}}=B\hat{z}$) defines as the thermal average of the corresponding components 
and it's square of the momentum transfer due to the HQ scattering process with thermal particles. 
The HQ drag force components can be defined as,
\begin{align}\label{1.8}
 &A_{\parallel}=<<q_z>>, &  A_{\perp}=\frac{1}{\sqrt{2}}<<q_{\perp}>>,
\end{align}
 where longitudinal and transverse components quantitatively measure the anisotropy in the drag force 
 in the presence of the magnetic field. Here, $q_{\perp}=\mid{\bf{q}}_{\perp}\mid$ is the transverse component 
 of the momentum transfer. Similarly, the components of the HQ momentum diffusion coefficients in the presence of the magnetic 
 field within the static limit can be defined as, 
 \begin{align}\label{1.10}
 &K_{\parallel}=<<q^2_z>>, &  K_{\perp}=\frac{1}{2}<<q_{\perp}^2>>.
\end{align}
Employing the fluctuation-dissipation theorem, we can define the longitudinal 
 and transverse drag coefficients respectively as,
 \begin{align}\label{1.9}
 &\eta_{\parallel}=\dfrac{K_{\parallel}}{2TM_{HQ}}, &  \eta_{\perp}=\dfrac{K_{\perp}}{2TM_{HQ}}.
\end{align}
We intend to compute the HQ drag force components and diffusion coefficients in the longitudinal and transverse 
directions considering the thermal medium effects of the magnetized QGP in the current analysis. The EoS effects can enter 
through both the parton momentum distribution function and the Debye screening mass (effective coupling) 
while defining the scattering amplitude of the interaction. Proper modeling of the hot magnetized QGP 
is essential to incorporate the effects of QCD medium interactions.

\subsubsection*{Modeling of hot magnetized QGP}

The effective modeling of the QGP by encoding the thermal QCD medium effects have been studied 
in several means such as self-consistent quasiparticle model~\cite{Bannur:2006js}, models based on the
Gribov-Zwanziger quantization~\cite{Su:2014rma,zwig,Bandyopadhyay:2015wua}, Nambu-Jona-Lasinio (NJL)
and Polyakov-loop-extended NJL based quasiparticle models~\cite{Dumitru}, effective mass quasiparticle model with Polyakov 
loop~\cite{D'Elia:97} and effective fugacity quasiparticle model (EQPM)~\cite{Chandra:2011en,Chandra:2007ca}. 
The present analysis is based on the EQPM in which the realistic EoS can be interpreted 
in terms of  temperature dependent quasigluon and quasiquark/antiquark fugacities, $z_{g}$ and $z_{q}$, respectively.
We consider the recent $(2+1)-$flavor lattice QCD EoS (LEoS)~\cite{Cheng:2007jq} from 
the lattice QCD simulations. Within the framework of the EQPM, the thermal QCD medium 
constitutes of effective gluonic sector and matter sector (light quarks).   
We have studied the EQPM for the magnetized QGP in the Ref~\cite{Kurian:2017yxj}.  
The EQPM quark distribution function in the presence 
of the magnetic field (${\bf{B}}=B\hat{z}$) has the following form,
\begin{equation}\label{2.1}
{f}^l_{q}=\dfrac{z_{q}\exp{(-\beta E^l_{k})}}{1+ z_{q}\exp{(-\beta E^l_{k})}}.
\end{equation}  
Here, $E^l_{k}=\sqrt{k_{z}^{2}+m^{2}+2l\mid q_feB\mid}$ is the Landau 
energy eigenvalue in which $l=0,1,2,..$ is the order of the Landau levels. 
Note that the effective fugacity parameter for quark and antiquark
is same, $i.e$.,$z_q=z_{\bar{q}}$ and hence the momentum distribution of quarks and 
antiquarks is identical (in the case of vanishing chemical potential). The dispersion relation 
of electrically chargeless gluon remain intact in the magnetic field and the
distribution function has the form,
\begin{equation}\label{2.2}
f_{g}=\dfrac{z_{g}\exp{(-\beta\mid{\bf{k}}\mid)}}
{1- z_{g}\exp{( -\beta\mid{\bf{k}}\mid})}.
\end{equation}
We choose the units $k_{B}=1$, $c=1$, 
$\hbar=1$ and $\beta=\dfrac{1}{T}$.
The physical significance of the effective fugacity parameter can be interpreted from the 
non-trivial dispersion relation which encodes the quasiparton collective 
excitations and takes the forms,
\begin{equation}\label{2.3}
\omega^l_{q}=\sqrt{k_z^{2}+m^{2}+2l\mid q_feB\mid}+T^{2}\partial_{T} \ln(z_{q}),
\end{equation}
and
\begin{equation}\label{2.4}
\omega_{g}=\mid{\bf{k}}\mid+T^{2}\partial_{T} \ln(z_{g}),
\end{equation}
for quarks and gluons, respectively. The effective Boltzmann equation and mean field terms 
are well investigated within the framework of the EQPM both in the presence and 
absence of magnetic field~\cite{Kurian:2018qwb,Mitra:2018akk}.

It is important to note that the EQPM is motivated from the charge renormalization in the 
QCD medium as the effective mass models are based on the mass renormalization. 
Both the dispersion relation and effective coupling are sensitive to the magnetic field in the medium.
The magnetic field dependence on the temperature behaviour of the effective coupling 
can be estimated from of the Debye screening mass of the magnetized QGP. 
The Debye screening mass in the QGP 
can be defined in terms of the EQPM momentum distribution function as~\cite{Kurian:2017yxj},
\begin{equation}\label{2.5}
m_{D}^{2}=-4\pi\alpha_{s}\int{d\Upsilon
\dfrac{d}{d{\bf{k}}}(2N_{c}{f}_{g}+ 2N_{f}{f}^l_{q})},
\end{equation}
where the integration phase factor $d\Upsilon=\frac{d^3{\bf{k}}}{(2\pi)^3}$ for gluons and 
$d\Upsilon=\frac{\mid q_feB\mid}{2\pi}\sum_{l=0}^{\infty}{\frac{dk_{z}}{2\pi}}(2-\delta_{0l})$ 
for quarks in the presence of the magnetic field. Here, $\alpha_{s}(T)$ is the QCD 
running coupling constant at finite temperature~\cite{Laine:2005ai}. The Eq.~(\ref{2.5}) 
can be solved separately for gluonic and light quark sector and takes the following form,
\begin{align}\label{2.6}
m_{D}^{2}&= \dfrac{24\alpha_{s}T^{2}}{\pi}PolyLog[2,z_{g}]\nonumber\\
&+\dfrac{4\alpha_{s}}{T}\sum_f\dfrac{\mid q_feB\mid}{\pi}
\int_{0}^{\infty}{\sum_{l=0}^{\infty}dk_z(2-\delta_{l0}){f}^l_q(1-{f}^l_q)}.
\end{align}
The term expressed in terms of $PolyLog$
function over the effective gluon fugacity parameter in the Eq.~(\ref{2.6}) represents the 
gluonic contribution to the screening mass. Whereas the second term  constitutes the 
contributions of quarks incorporating the HLL effects. For the system of ultra-relativistic 
non-interacting quarks and gluons (ideal EoS), $z_{q,g}=1$, the Debye mass takes form,
\begin{align}\label{2.7}
{m^{2}_{D}}^{ideal}&=4\pi\alpha_{s}(T)\bigg[T^{2}+\sum_f{\Lambda_f(\mid q_feB\mid, T)}\bigg],
\end{align}
in which the quark (with flavour $f$) contribution in the ideal case ($z_q=1$) can be defined as,
\begin{equation}\label{2.8}
\Lambda_f=\frac{1}{T}\frac{\mid q_feB\mid}{\pi^2}
\int_{0}^{\infty}{\sum_{l=0}^{\infty}dk_z(2-\delta_{l0}){n}^l_q(1-{n}^l_q)},
\end{equation}
where ${n}^l_{q}=\frac{1}{\exp{(\beta E^l_k)}+ 1}$ is the Fermi-Dirac distribution function. 
Defining the effective running coupling constant $\alpha^l_{eff}(T,z_q,z_g,\mid eB\mid)$ as, 
$m^{2}_{D}=\dfrac{\alpha^l_{eff}}{\alpha_s}{m^{2}_{D}}^{ideal}$
and employing Eq.~(\ref{2.6}) and Eq.~(\ref{2.7}), we obtain,
\begin{align}\label{2.9}
\dfrac{\alpha^l_{eff}}{\alpha_{s}}&=
\dfrac{ {6T^{2}}
PolyLog[2,z_{g}]}
{{\pi^{2}}\bigg( T^{2}+\sum_f{\Lambda_f(\mid q_feB\mid, T)}\bigg)}\nonumber\\&
+ \dfrac{\sum_f \mid q_feB\mid
\int_{0}^{\infty}{\sum_{l=0}^{\infty}dk_z(2-\delta_{l0}){f}^l_q(1-{f}^l_q)}}{{T\pi^{2}}\bigg( T^{2}
+\sum_f{\Lambda_f(\mid q_feB\mid, T)\bigg)}}.
\end{align}
Note that the hot QCD medium effects are entering through the quasiparton momentum distribution 
functions and the scattering amplitude. The effective coupling constant, which is a dynamical input 
of the HQ transport, is incorporated through the HQ scattering with thermal gluons and quarks.  
We use these concepts in the estimation of HQ drag and diffusion in the presence of 
the magnetic field in the next sections.  

\subsubsection*{Thermal gluon contribution to HQ drag and diffusion}

The gluon contribution to the HQ transport coefficients is incorporated through the HQ scattering 
with thermal gluons in the hot medium. The  $2\leftrightarrow 2$ HQ-gluon scattering is dominated 
by t-channel gluon exchange~\cite{Moore:2004tg} and the matrix element $\mathcal{M}_{HQ,g}$ 
in the static limit has the following form~\cite{Fukushima:2015wck},
\begin{align}\label{3.1}
\mathcal{M}_{HQ,g}^{bc}=& -i8 \pi~\alpha_{eff} M_{HQ}~f^{abc}{t_R}^a \dfrac{1}{\big(q^2+s(q_{\perp})\big)}\nonumber\\
&\times\Big(\mid{\bf{k}}\mid +\mid{\bf{k}^{'}}\mid\Big)\epsilon.\bar{\epsilon}^{*}.
\end{align}
Here, $q\equiv\mid{\bf q}\mid$ and $\alpha_{eff}$ is the effective coupling in medium. Here, 
$(b,\epsilon)$ and $(c,\bar{\epsilon})$ are the representation of color and polarization of the incoming 
and outgoing gluons. The quantity $f^{abc}$ is the structure constant and $t^a_R$ is the generator of 
the group $SU(N_c)$. The quantity $s(q_{\perp})$ in the strong magnetic field is defined as,
\begin{align}\label{3.22}
s(q_{\perp})=4\pi\alpha_{eff}~\sum_f{\Lambda_f}~\exp{\Big(\dfrac{-q_{\perp}^2}{2\mid q_feB\mid}\Big)},
\end{align}
where $\Lambda_f$ is defined in the Eq.~(\ref{2.8}) such that $s(q_{\perp}=0)$ gives the 
quark contribution to the Debye mass in the magnetized medium.
In the LLL approximation, the Eq.~(\ref{3.22}) reduced to the form 
$s(q_{\perp})=\alpha_s\sum_f\frac{\mid q_feB\mid}
{\pi}~\exp{\Big(\frac{-q_{\perp}^2}{2\mid q_feB\mid}\Big)}$ for the non-interacting case. 
We can compute $\mid\mathcal{M}_{HQ,g}\mid^2$ from Eq.~(\ref{3.1}) in the static limit ($\mid{\bf{k}}\mid=\mid{\bf{k}}^{'}\mid$) 
by employing the polarization sum, $\sum_{\epsilon,\bar{\epsilon}^{*}}\mid\epsilon.\bar{\epsilon}^{*}\mid^2=
1+\cos^2\theta_{{{\bf k}}{{\bf k}}^{'}}$, where $\theta_{{{\bf k}}{{\bf k}}^{'}}$ is the angle between ${\bf k}$ 
and ${{\bf k}}^{'}$ and the color summation, $\sum{f^{abc}f^{a^{'}bc}t_R^at_R^{a^{'}}}=N_cC_R^{HQ}I$, 
in which $C_R^{HQ}$ is the color Casimir of the HQ. Incorporating all these arguments, 
we can define the gluon contribution to the longitudinal HQ drag component 
by substituting the Eq.~(\ref{3.1}) in the Eq.~(\ref{1.6}) as,
\begin{align}\label{3.2}
A^{gluon}_{\parallel}&=\dfrac{4}{\pi}\alpha_{eff}^2N_cC_R^{HQ}\dfrac{1}{d_{HQ}}\int_{0}^{\infty}{dq}~q^2\dfrac{1}{\big(q^2+s(q_{\perp})\big)^2}\nonumber\\
&\times\int_{0}^{\infty}{d\mid{\bf k}\mid}\mid{{\bf k}}\mid^2(1+\cos^2\theta_{{{\bf k}}{{\bf k}}^{'}})\delta(\mid{{\bf k}}\mid-\mid{{\bf k}}-{\bf q}\mid)\nonumber\\&
\times f_g(1+f_g)~q_z.
\end{align}
Following the prescriptions in~\cite{Fukushima:2015wck}, we can further simplify the 
Eq.~(\ref{3.2}) by considering the rotational symmetry such that $q_z^2=\frac{q^2}{3}$ 
and employing $\delta(\mid{{\bf k}}\mid-\mid{{\bf k}}-{{\bf q}}\mid)=
q^{-1}\delta\big(\cos\theta_{{\bf k}{\bf q}}-\frac{q}{2\mid{\bf k}\mid}\big)\Theta(\mid{\bf k}\mid-\frac{q}{2})$ 
and $\cos\theta_{{{\bf k}}{{\bf k}}^{'}}=1-\frac{q^2}{2\mid{\bf k}\mid^2}$. Finally, Eq.~(\ref{3.2}) becomes,
\begin{align}\label{3.3}
A^{gluon}_{\parallel}&=\dfrac{4}{\sqrt{3}\pi}\alpha_{eff}^2N_cC_R^{HQ}\dfrac{1}{d_{HQ}}
\int_{0}^{\infty}{dq}~\dfrac{q^2}{\big(q^2+s(q_{\perp})\big)^2}\nonumber\\
&\times\int_{q/2}^{\infty}{d\mid{\bf k}\mid}\mid{\bf k}\mid^2\Big[1+\Big(1-\dfrac{q^2}{2\mid{\bf k}\mid^2}\Big)^2\Big] f_g(1+f_g).
\end{align}
Similarly, we can calculate the quark contribution to longitudinal HQ momentum diffusion 
by substituting Eq.~(\ref{3.1}) in Eq.~(\ref{1.7}) and has the following form,
\begin{align}\label{3.31}
K^{gluon}_{\parallel}&=\dfrac{4}{3\pi}\alpha_{eff}^2N_cC_R^{HQ}\dfrac{1}{d_{HQ}}\int_{0}^{\infty}{dq}~q^2
\dfrac{1}{\big(q^2+s(q_{\perp})\big)^2}\nonumber\\
&\times\int_{0}^{\infty}{d\mid{\bf k}\mid}\mid{{\bf k}}\mid^2(1+\cos^2\theta_{{{\bf k}}{{\bf k}}^{'}})
\delta(\mid{{\bf k}}\mid-\mid{{\bf k}}-{\bf q}\mid)\nonumber\\&
\times f_g(1+f_g)~q^2.
\end{align}
Performing the $q$ integral and dropping out the terms sub-leading to $m_D/T$ by assuming 
the contribution from hard thermal gluons with $\mid{\bf k}\mid\gtrsim T$ as in the described 
in the Refs.~\cite{Moore:2004tg,CaronHuot:2007gq}, the leading order $K_{\parallel}^{gluon}$ 
in the presence of magnetic field takes the following form,
\begin{align}\label{3.4}
K^{gluon}_{\parallel}=&\dfrac{8\pi}{9}\alpha_{eff}^2N_cC_R^{HQ}\dfrac{1}{d_{HQ}}T^3\bigg[\log\Big(\dfrac{2T}{m_D}\Big)+2~\xi\bigg],
\end{align}
where $\xi=\frac{1}{2}-\gamma_E+\frac{\zeta^{'}(2)}{\zeta(2)}$, in which $\gamma_E$ and $\zeta$ are 
the Euler's constant and Riemann zeta function, respectively. The effects of hot QCD medium interactions 
and HLLs in the temperature behaviour of the HQ transport coefficients are entering through momentum 
distribution function and effective coupling in the medium.  

For the non-interacting case ($z_{q,g}=1$), for the LLL quarks we have,  
$\log\big(\frac{2T}{m_D}\big)\approx \Big[\log(\frac{1}{\alpha_s}\big)-\log(\frac{\sum_f\mid q_feB\mid}{T^2\pi})\Big]$. 
The form of LLL quark contribution to the longitudinal momentum diffusion in the ideal EoS is consistent 
with the observations in the recent work~\cite{Fukushima:2015wck}. Since the gluonic dynamics are 
not directly affected by the magnetic field, the gluon contribution to the HQ drag force and momentum diffusion
are isotropic in nature as,  
\begin{align}\label{3.5}
 &A^{gluon}_{\parallel}=A^{gluon}_{\perp}, &K^{gluon}_{\parallel}=K^{gluon}_{\perp}.   
\end{align} 
Whereas the quark contributions to the transport coefficients are highly anisotropic and are discussed in the next section.

\subsection{Leading order quark contributions to HQ drag and diffusion}

To incorporate the quark contribution to the HQ dynamics, we consider the process 
$HQ(p)+l(k)\rightarrow HQ(p^{'})+l(k^{'})$, where $l$ represents the light quarks in the thermal medium. 
The quark dynamics is significantly affected by the strong magnetic field and follow the $1+1-$dimensional 
Landau level dynamics. The HQ motion is subjected to the scattering with the quarks both in the direction 
parallel and perpendicular to the magnetic field. 

\subsubsection{HQ longitudinal and transverse drag}

\begin{figure*}
 \centering
 \subfloat{\includegraphics[scale=0.4]{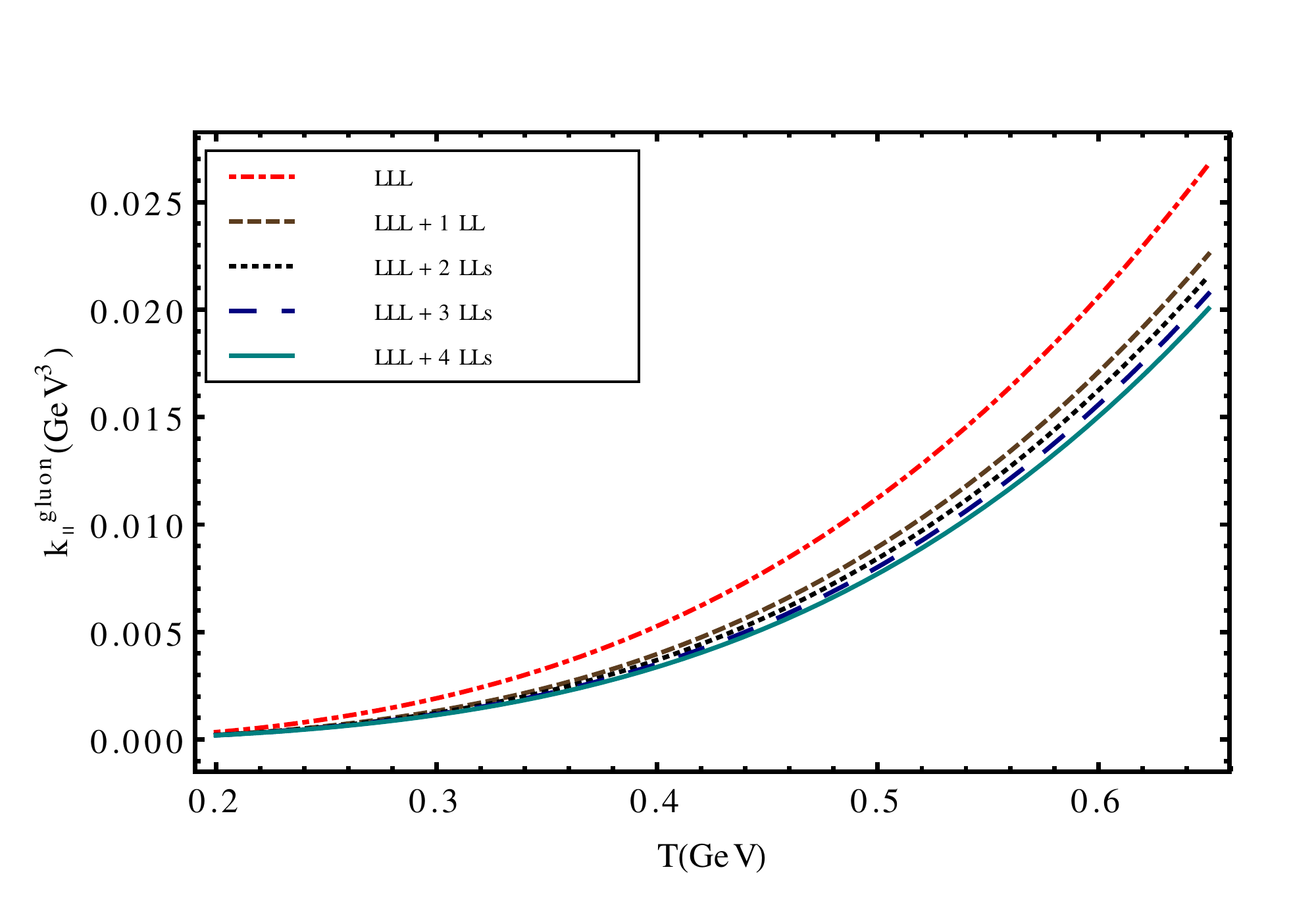}}
 \hspace{1 cm}
 \subfloat{\includegraphics[scale=0.4]{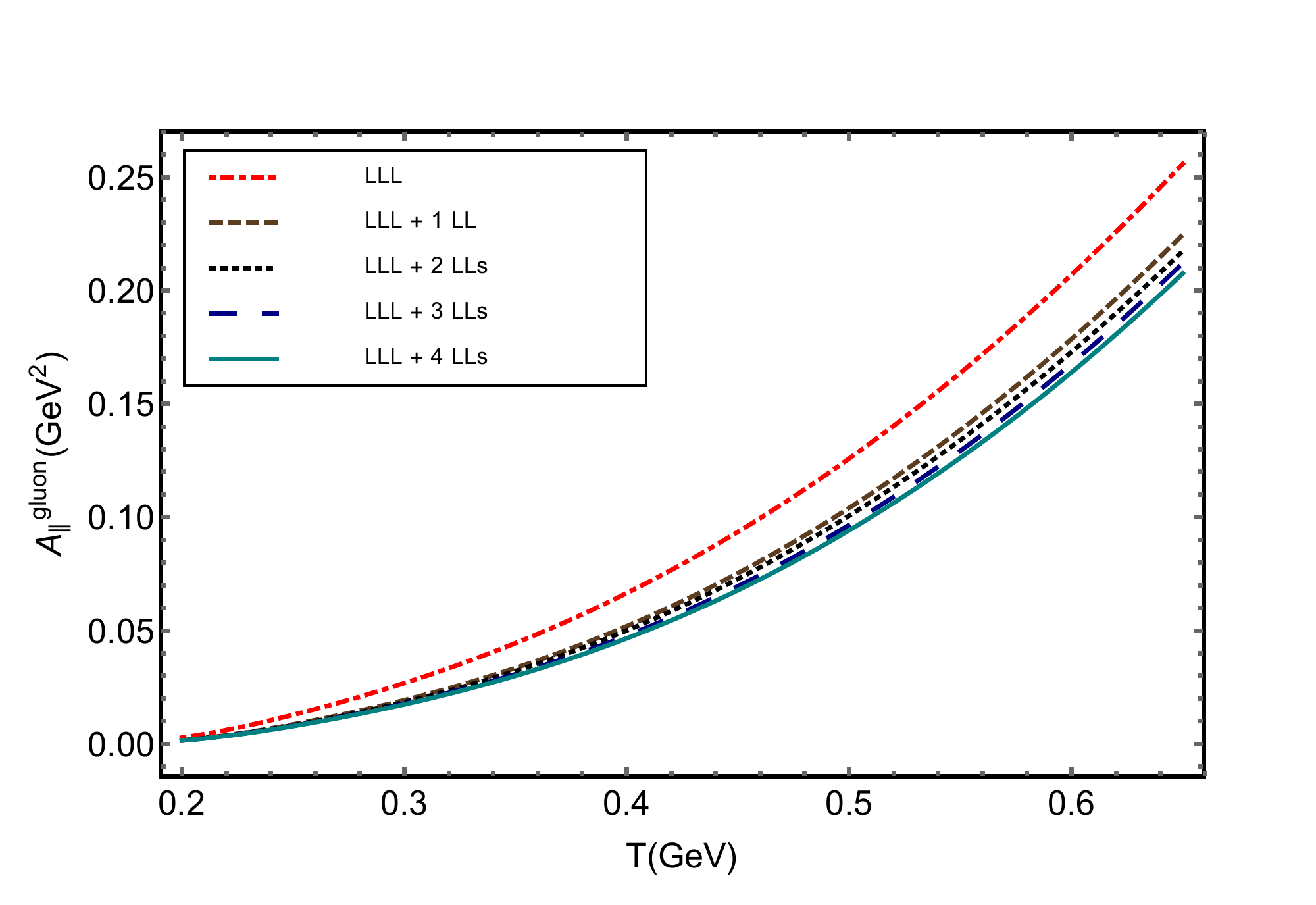}}
\caption{(Color online) The HLLs effect in the temperature behaviour of the $K^{gluon}_{\parallel}$ (left panel) and $A^{gluon}_{\parallel}$(right panel)in the static limit at $\mid eB\mid=15m_{\pi}^2$.}
\label{f1}
\end{figure*} 
The longitudinal and transverse components of HQ drag for the HQ-quark scattering process can 
be described in the static limit by following the same prescription as that of the gluonic case 
as in Eq.~(\ref{1.6}) and takes the forms as,
\begin{align}\label{4.1}
A^{quark}_{\parallel}=\int{d^3{\bf q}\dfrac{d\Gamma({\bf q})}{d^3{\bf q}}q_z},
\end{align}
and
\begin{align}\label{4.2}
A^{quark}_{\perp}=\frac{1}{\sqrt{2}}\int{d^3{\bf q}\dfrac{d\Gamma({\bf q})}{d^3{\bf q}}q_{\perp}}.
\end{align}
The quantity $\frac{d\Gamma({\bf q})}{d^3{\bf q}}$ defines the HQ scattering rate with quarks per 
unit volume of momentum transfer via one gluon exchange and has the following form,
\begin{align}\label{4.3}
\dfrac{d\Gamma}{d^3{\bf q}}=&\dfrac{1}{d_{HQ}}\dfrac{1}{2M_{HQ}}\dfrac{1}{(2\pi)^32E_q}\int{\dfrac{d\Upsilon}{E_k^l}}\int{\dfrac{d\Upsilon^{'}}{E_{k^{'}}^l}}\mid\mathcal{M}_{HQ,q}\mid^2\nonumber\\
&\times(2\pi)^2\delta^2(p+k-p^{'}-k^{'})f^l_q(k_z)
\Big(1-f^l_q(k^{'}_z)\Big),
\end{align}
with $f^l_q(k_z)$ and $E_k^l$ are the EQPM distribution function and Landau level energy eigenvalue 
of the quark, respectively. Here, $\mid\mathcal{M}_{HQ,q}\mid$ is the matrix element of the HQ scattering with 
the thermal quarks. In the recent work~\cite{Fukushima:2015wck}, the authors have estimated the 
HQ-quark scattering rate from the retarded gluon correlator by employing the real time 
Schwinger-Keldysh formalism and has the following form,
\begin{align}\label{4.4}
\dfrac{d\Gamma}{d^3{\bf q}}=\dfrac{\alpha_{eff}T}{\pi} N_cC_{R}^{HQ}\dfrac{s(q_{\perp})}{\Big(q^2+s(q_{\perp})\Big)^2} \delta(q_z), 
\end{align}
in which $\alpha_{eff}$ is the effective coupling as described in the Eq.~(\ref{2.9}). 
The contribution from the quarks with LLL and the HLLs next to $l=0$ state constitute the leading order quark 
contribution to the HQ drag and diffusion in the strong magnetic field background. We conclude 
from Eq.~(\ref{4.1}) and Eq.~(\ref{4.4}) that in the strong magnetic field, the quark contribution to the 
longitudinal drag component goes to zero. By substituting Eq.~(\ref{4.4}) in Eq.~(\ref{4.2}), the leading order 
transverse HQ drag takes the following form,
\begin{align}\label{4.6}
A_{\perp}^{quark}=&~2~T\alpha^2_{eff}~N_cC_R^{HQ}~\dfrac{\sqrt{\mid eB\mid}}{\pi}\nonumber\\
&\times\int_{0}^{\infty}{dx\dfrac{\sqrt{x}~N(T,x)}{\Big(x+2\dfrac{\alpha_{eff}}{\pi}N(T,x)\Big)^2}},
\end{align}
where $x=\frac{q^2_{\perp}}{2\mid eB\mid}$ and $N(T,x)$ is defined as,
\begin{equation}\label{4.7}
N=\dfrac{1}{T}\sum_f\mid q_f\mid e^{-\frac{x}{\mid q_f\mid}}
\int_{0}^{\infty}{\sum_{l=0}^{\infty}dk_z(2-\delta_{l0}){n}^l_q(1-{n}^l_q)}.
\end{equation}
The thermal quark contribution to the HQ drag is in the transverse direction and have a 
dependence on the HLLs and non-ideal EoS.

\begin{figure*}
 \centering
 \subfloat{\includegraphics[scale=0.4]{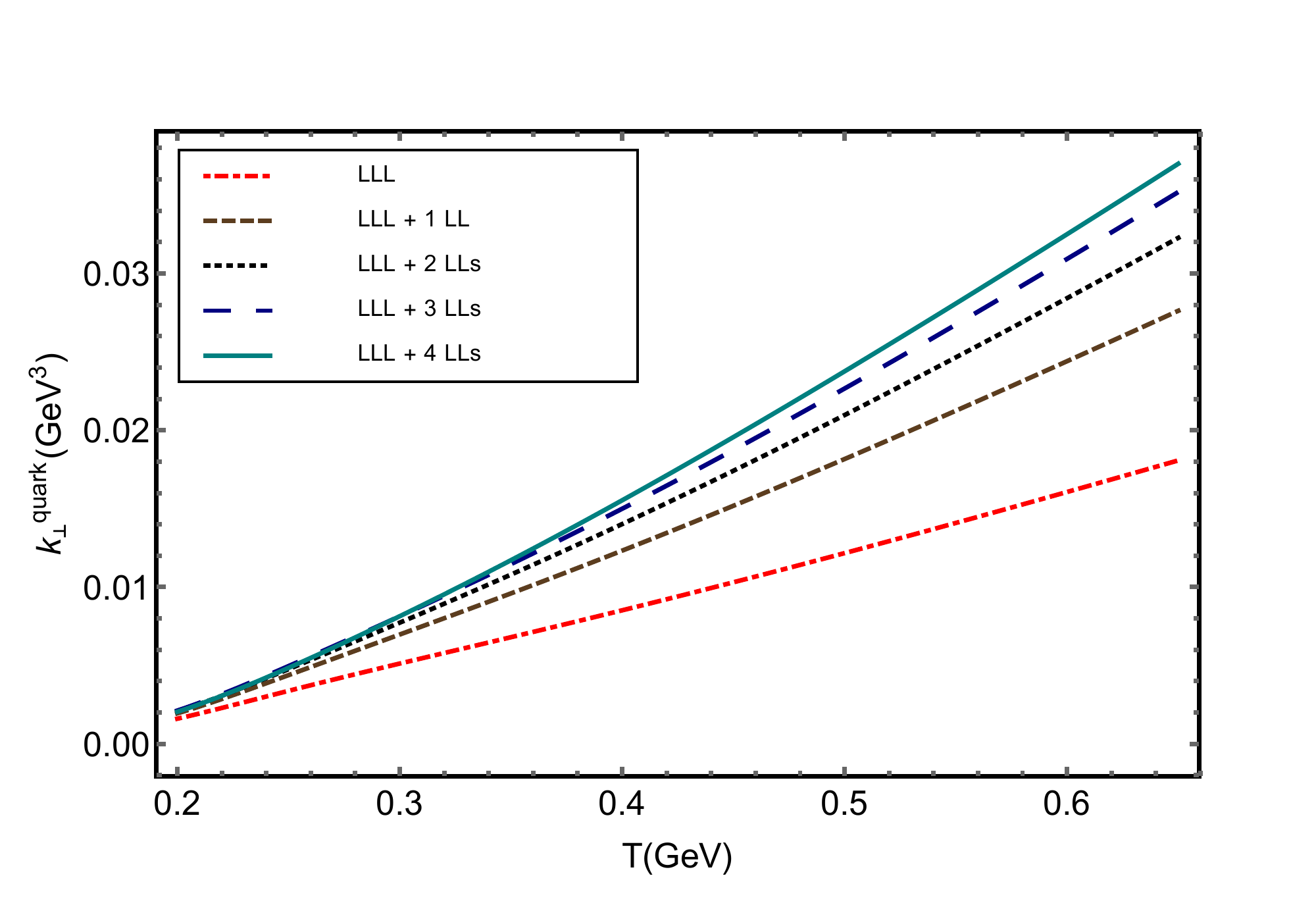}}
 \hspace{1.15 cm}
 \subfloat{\includegraphics[scale=0.4]{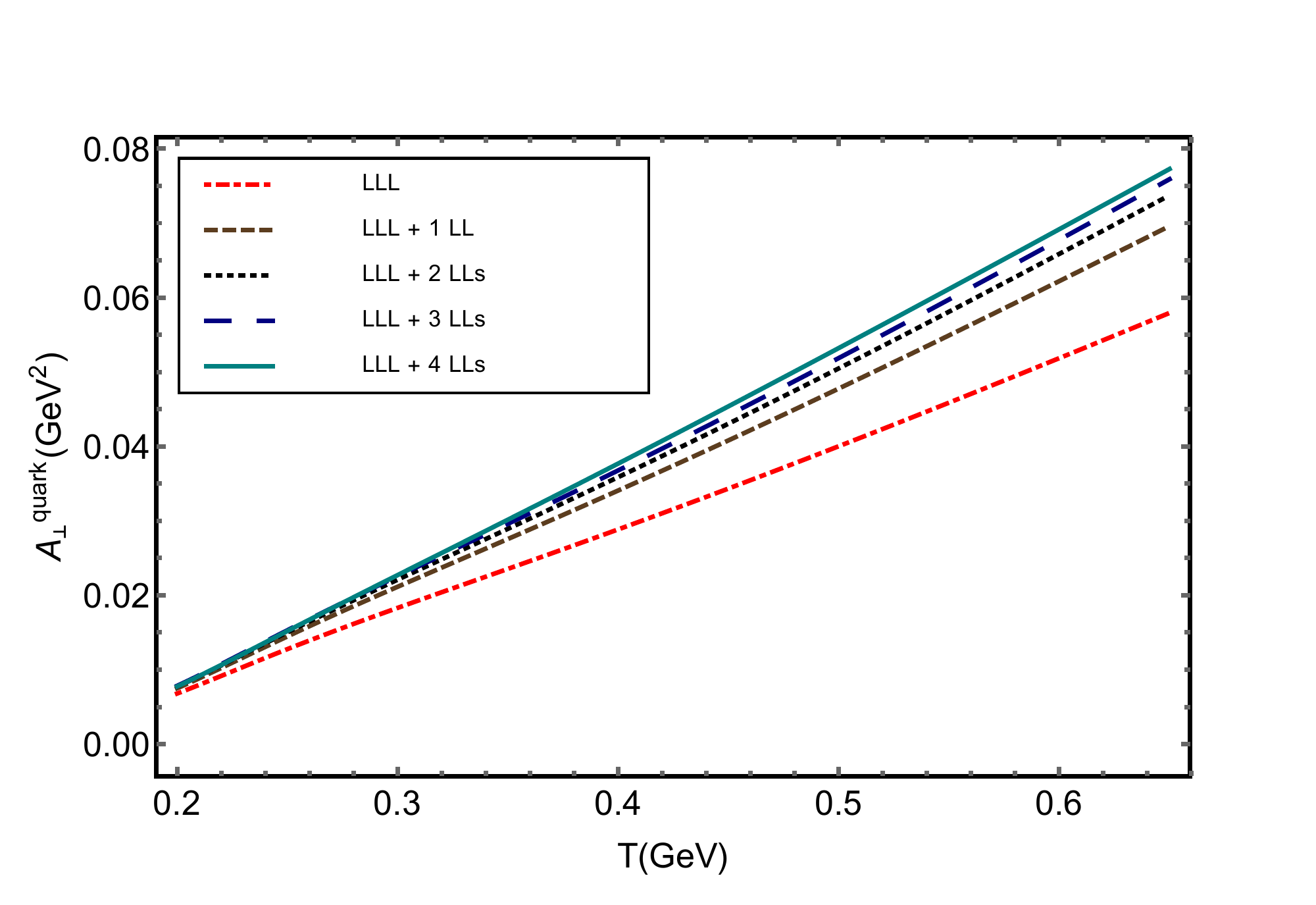}}
\caption{(Color online) The HLLs effect in the temperature dependence of the $K^{quark}_{\perp}$ (left panel) and $A^{quark}_{\perp}$(right panel) at $\mid eB\mid=15m_{\pi}^2$.}
\label{f2}
\end{figure*}
\begin{figure*}
 \centering
 \subfloat{\includegraphics[scale=0.475]{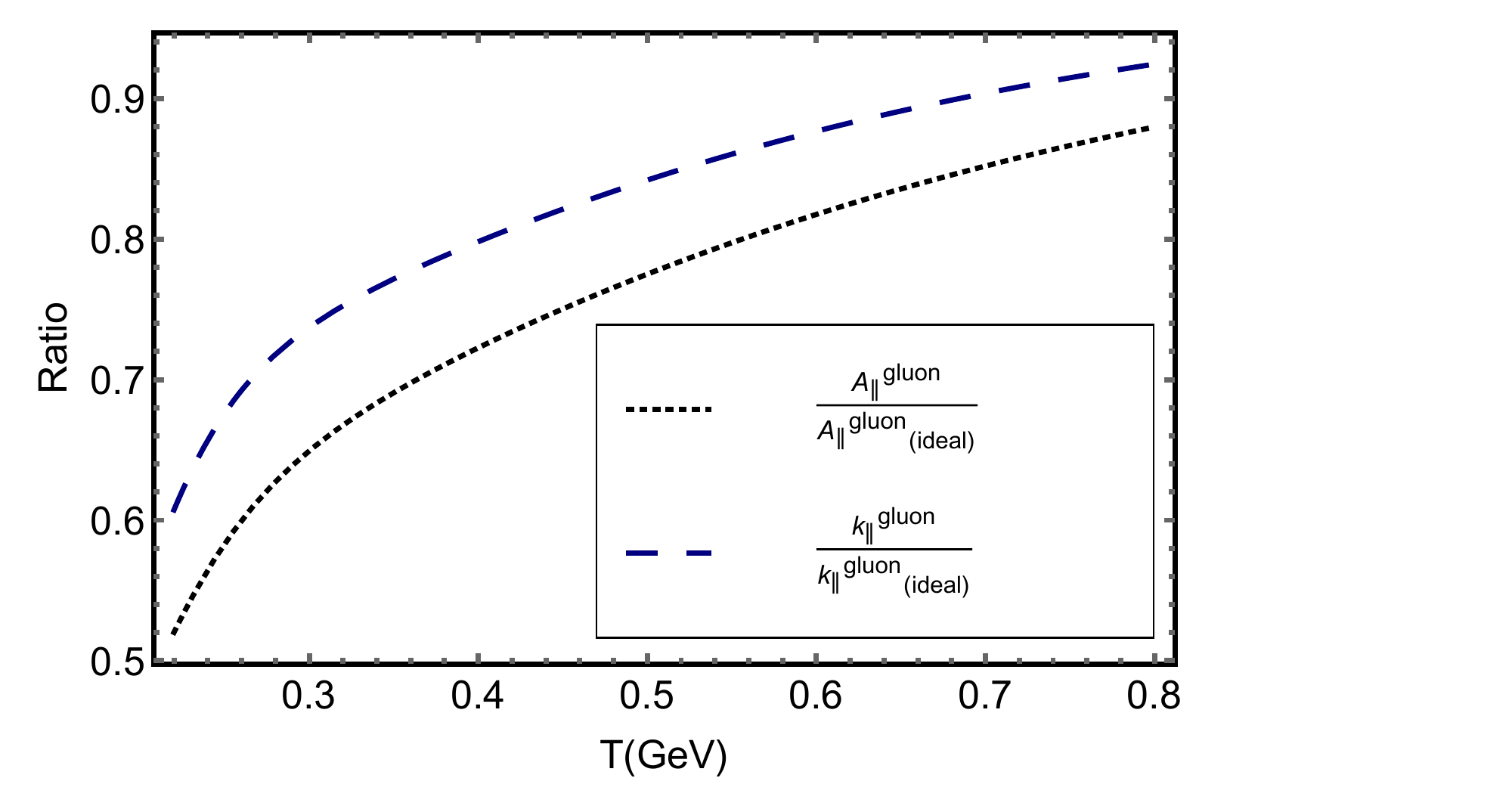}}
 \subfloat{\includegraphics[scale=0.475]{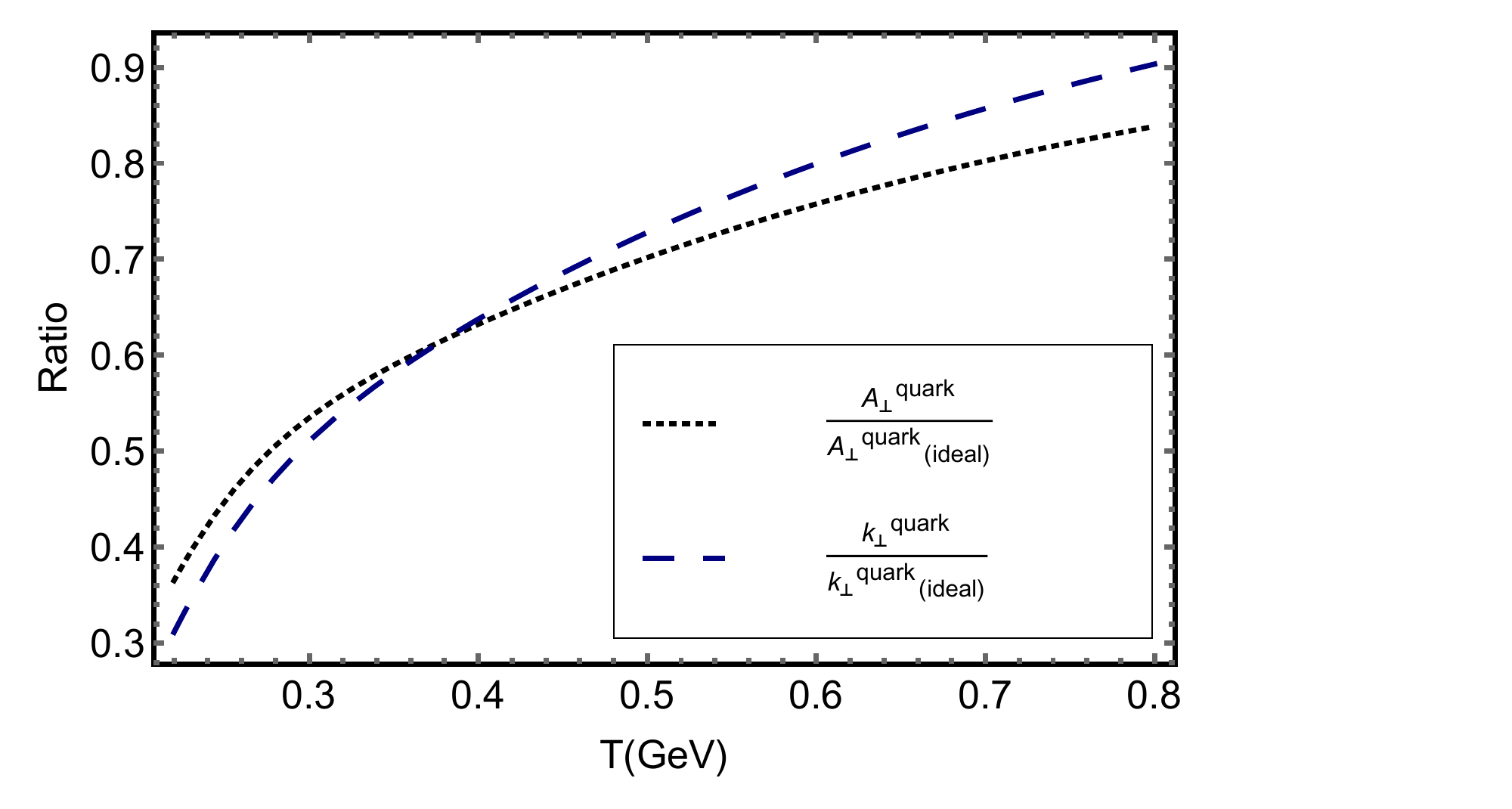}}
\caption{(Color online) The EoS dependence on the longitudinal (left panel) and transverse (right panel) 
HQ momentum diffusion and drag at $\mid eB\mid=15m_{\pi}^2$.}
\label{f3}
\end{figure*}
\subsubsection{HQ diffusion}

Similar to the HQ momentum described in Eq.~(\ref{1.7}) due to the HQ-gluon scattering, the quark 
contribution to the longitudinal and diffusion  coefficients take the forms as follows,
\begin{align}\label{5.1}
K^{quark}_{\parallel}=\int{d^3{\bf q}\dfrac{d\Gamma({\bf q})}{d^3{\bf q}}q^2_z},
\end{align}
and
\begin{align}\label{5.2}
K^{quark}_{\perp}=\dfrac{1}{2}\int{d^3{\bf q}\dfrac{d\Gamma({\bf q})}{d^3{\bf q}}q_{\perp}^2}.
\end{align}
By substituting the definition of HQ-quark scattering rate in Eq.~(\ref{5.2}), we obtain the leading order quark 
contribution to the transverse momentum diffusion as follows,
\begin{align}\label{5.3}
K_{\perp}^{quark}=&4~T~\alpha^2_{eff}~N_cC_R^{HQ}~\dfrac{{\mid eB\mid}}{2\pi}\nonumber\\
&\times\int_{0}^{\infty}{dx\dfrac{{x}~N(T,x)}{\Big(x+2\dfrac{\alpha_{eff}}{\pi}N(T,x)\Big)^2}},
\end{align}
where $N(T,x)$ is defined in the Eq.~(\ref{4.7}). The longitudinal component of HQ diffusion due to 
scattering with quarks vanishes in the strong magnetic field and can be understood from the 
Eq.~(\ref{4.4}) and Eq.~(\ref{5.1}). Similar to the drag force, the quark contribution to the 
HQ momentum diffusion in the magnetic field is anisotropic in nature. In the ideal EoS ($z_{q,g}=1$), our results 
for LLL can be reduced to that in the Ref.~\cite{Fukushima:2015wck}.


\section{Results and Discussion}
\begin{figure}[h]
  \subfloat{\includegraphics[scale=0.4]{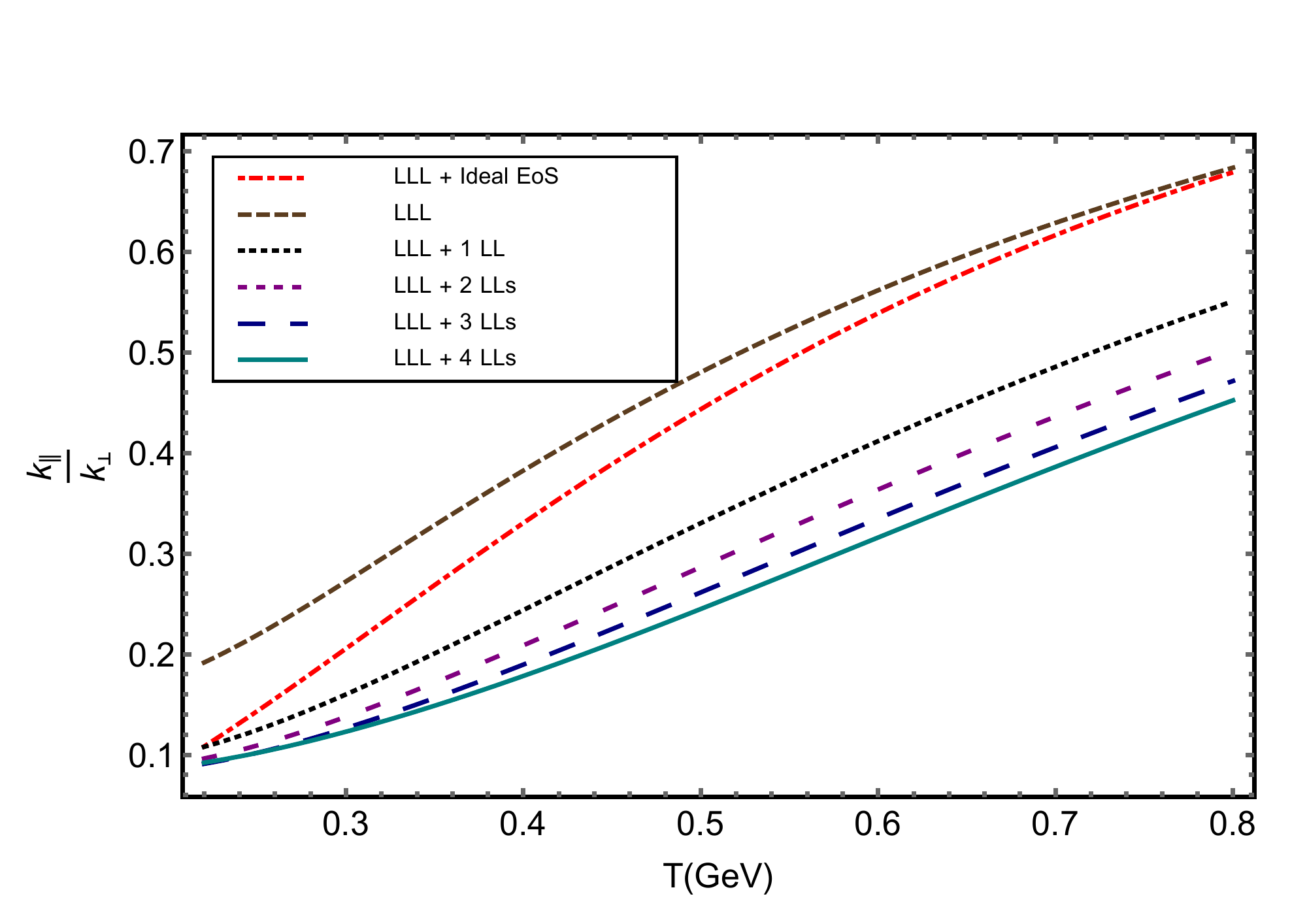}}
\caption{(Color online) The effects of HLLs and EoS in the magnetic field induced 
anisotropy in the HQ momentum diffusion.  }
\label{f4}
\end{figure}

We initiate the discussion with the effects of HLLs on the temperature dependence of the HQ 
momentum diffusion and drag in the magnetized QGP. Since the thermal quarks and gluons 
follow different dynamics in the magnetic field, we have considered HQ scattering with quarks 
and gluons separately, along the direction parallel and perpendicular to the magnetic field. 
The gluonic contribution to the HQ drag force and momentum diffusion is incorporated to the 
HQ-thermal gluon scattering process. The temperature behaviour of the $K^{gluon}_{\parallel}$ 
and $A^{gluon}_{\parallel}$ at $\mid eB\mid=15m_{\pi}^2$ is depicted in the Fig.\ref{f1}. The 
HLLs effects to the gluonic contributions can enter through the Debye screening mass as 
described in the Eq.~(\ref{3.3}) and Eq.~(\ref{3.4}). For the numerical estimation of the effect 
of HLLs, we plotted with different Landau levels. We truncate the Landau levels at $l=4$ 
and the HLL contributions beyond $l=4$ seems to be negligible in the chosen range of temperature. 
We observe that the HLL contribution quantitatively reduces the longitudinal HQ momentum 
diffusion and drag, and the effect is more pronounced in the high temperature regime. The gluons 
are not directly coupled to the magnetic field and follow $(1+3)-$dimensional dynamics. Hence, 
the gluonic contribution to the HQ drag and diffusion is isotropic in nature. 

We have incorporated the $(1+1)-$dimensional Landau level kinematics for quarks while estimating the 
quark contribution to the HQ transport from the HQ-thermal quark scattering in the 
presence of the magnetic field. The effect of HLLs on the temperature behaviour of the 
$K^{quark}_{\perp}$ and $A^{quark}_{\perp}$ are  shown in the Fig.\ref{f2}. We 
observe that HLL corrections enhance the quark contribution to the transverse components 
of the drag force and diffusion coefficient. The HLLs effect is quite significant in the higher 
temperature regime. The quark contribution in the longitudinal direction vanishes in the 
leading order and can be understood from Eq.~(\ref{4.1}), Eq.~(\ref{4.4}) and Eq.~(\ref{5.1}). 
Our observation on the anisotropic nature of the HQ drag and diffusion is qualitatively 
consistent with the results in the Ref.~\cite{Fukushima:2015wck}.

The thermal medium dependence of the HQ drag force and momentum diffusion is governed by the quasiparticle momentum 
distribution function and effective coupling. 
We plotted the EoS dependence of the drag force and diffusion coefficient for both gluonic 
and quark contributions at $\mid eB\mid=15m_{\pi}^2$ in the Fig.\ref{f3}. The EoS dependence 
in HQ transport is more visible in the temperature regime near to the transition temperature, 
$T_c=0.17$ GeV.  Asymptotically, the ratio tends to unity, which implies that the quasipartons 
will behave like free particles at very high temperature regime. 

In Fig.\ref{f4}, we depicted the effects of HLLs and EoS in the anisotropy in the HQ momentum 
diffusion by estimating the temperature behaviour of the ratio $\dfrac{K_{\parallel}}{K_{\perp}}$. 
The quantities $K_{\parallel}$ and $K_{\perp}$ represent the total contribution from the quark 
and gluonic sector to the longitudinal and transverse diffusion coefficient in the presence of 
the magnetic field. We observe that the HLLs enhance the anisotropy in the HQ momentum 
diffusion, especially in the higher temperature region. For the LLL case, we have 
$K_{\perp}\gg K_{\parallel}$ in the temperature regime near to $T_c$. This observation 
for the LLL case is in line with the results of the recent work~\cite{Fukushima:2015wck}. 
The anisotropy in the HQ drag coefficients can be understood from the Eq.~(\ref{1.9}) in the hot QCD medium.

\section{Conclusion and Outlook}

In conclusion, we have computed the temperature dependence of the longitudinal and transverse 
components of the HQ drag force and momentum diffusion in the presence of magnetic field by 
considering the thermal gluonic and quark contributions, separately. We have employed the 
Fokker-Planck dynamics to describe HQ transport in the hot magnetized medium. The HQ drag 
and diffusion are influenced by the magnetic field and hot QCD medium. We observed that the 
inclusion of HLLs is essential for the description of drag force and momentum diffusion of the HQs at the higher 
temperature regime far from the transition temperature. Notably, the HLLs quantitatively suppresses 
the gluonic contribution to the HQ drag and diffusion, whereas the quark contribution to the 
transverse components gets enhanced at the high temperature regime. 
The magnetic fields effects are embedded through the Landau levels in the quasiquark momentum 
distribution functions and in the respective energy dispersion relations. On the other hand, the gluon
kinematics is coupled with the magnetic field through the effective coupling. Thermal medium effects 
are incorporated through the quasiparticle description of the magnetized medium. Furthermore, we have 
studied the anisotropy in the HQ momentum diffusion that induced from the $(1+1)-$dimensional Landau 
dynamics of the thermal quarks in the presence of the magnetic field. Finally, both the EoS and HLLs 
are seen to have a significant impact on the HQ momentum diffusion anisotropy in the magnetized medium.

HQ directed flow~\cite{Das:2016cwd} is considered as a sensitive probe to the creation and 
characterization of the magnetic field in the heavy-ion collisions. The anisotropic momentum diffusion 
and drag coefficients of the HQs due to the magnetic field can affect the HQ directed flow measured 
recently both at RHIC and LHC energies~\cite{Grosa:2018zix,Singha:2018cdj,Adam:2019wnk}. HQ elliptic 
flow is another observable which can be affected by this anisotropic HQ transport coefficients. The estimation of the HQ transport 
coefficients beyond the static limit for an arbitrary magnetic field is another important task.  
We intend to work on these  interesting aspects of the HQ dynamics in the near future.

\section*{acknowledgments}
We are highly thankful to Charles Gale for useful comments and suggestions. M.K. would like to acknowledge the 
hospitality of McGill University and the IIT Gandhinagar Overseas Research Experience
Fellowship to visit McGill University, where a part of this work is completed.
V.C. would like to acknowledge SERB for the Early 
Career Research Award (ECRA/2016), and DST, Govt. of India for INSPIRE-Faculty Fellowship 
(IFA-13/PH-55). S.K.D. acknowledges the support by the National Science Foundation of China 
(Grants No.11805087 and No. 11875153). We are indebted to the people of India for their 
generous support for the research in basic sciences. 



{}

\end{document}